\documentclass[aps,showkeys,superscriptaddress,notitlepage]{revtex4-1}
\usepackage{graphics,graphicx} 
\usepackage{geometry}
\usepackage{epstopdf}
\usepackage{natbib}
\begin{document}

\title
{Seasonal and geographical impact on  human resting periods}

\author{Daniel Monsivais}
\email[Corresponding author; ]{daniel.monsivais-velazquez@aalto.fi}
\author{Kunal Bhattacharya}
\author{Asim Ghosh}
\affiliation {Department of Computer Science, Aalto University School of Science, P.O. Box 15400, FI-00076 AALTO, Finland}
\author{Robin I.M. Dunbar}
\affiliation {Department of Experimental Psychology, University of Oxford, South Parks Rd, Oxford, OX1 3UD, United Kingdom}
\affiliation {Department of Computer Science, Aalto University School of Science, P.O. Box 15400, FI-00076 AALTO, Finland}
\author{Kimmo Kaski}
\affiliation {Department of Computer Science, Aalto University School of Science, P.O. Box 15400, FI-00076 AALTO, Finland}
\affiliation {Department of Experimental Psychology, University of Oxford, South Parks Rd, Oxford, OX1 3UD, United Kingdom}
\vspace*{0.2in}

\maketitle

\section*{Abstract}
We study the influence of seasonally and geographically related daily dynamics of daylight and ambient temperature on human resting or sleeping patterns using mobile phone data of a large number of individuals. We observe two daily inactivity periods in the people's aggregated mobile phone calling patterns and infer these to represent the resting times of the population. 
We find that the nocturnal resting period is strongly influenced by the length of daylight, and that its seasonal variation depends on the latitude, such that for people living in two different cities separated by eight latitudinal degrees, the difference in the resting period of people between the summer and winter in southern cities is almost twice that in the northern cities.
We also observe that the duration of the afternoon resting period is influenced by the temperature, and that there is a threshold from which this influence sets in. Finally, we observe that the yearly dynamics of the afternoon and nocturnal resting periods appear to be counterbalancing each other. This also lends support to the notion that the total daily resting time of people is more or less conserved across the year.

\section*{Introduction}
In modern societies, daily activities are entrained by at least two different diurnal rhythms, each one following a 24-hour clock with a different offset. The first one is marked by environmental and sun-based events that are subject to seasonal variations due to the yearly movement of earth around the sun. The second one follows a local civil time, where social and economical factors impose restrictions on the timing and routine of human activities \cite{aschoff1976human}. Regardless of the nature of the diurnal rhythm, modifications in it have a direct influence on various aspects of human life \cite{lange2010effects,souetre1989circadian}.

For human circadian rhythms, particularly for the sleep wake cycle (SWC), there are various biological \cite{czeisler1986bright}, sociological \cite{grandin2006social} and environmental \cite{roenneberg2007human} factors that influence the entrainment with the exogenous clocks, sometimes with undesirable effects on the mental and physical health of individuals  \cite{lewy2006circadian,lange2010effects,orzel2010consequences,evans2013health,moller2013effects,davies2014effect,santhi2016sex}. Due to its importance, the SWC has been studied in recent years from different perspectives, trying to understand and identify what are the processes and pacemakers governing its dynamics \cite{wright2001intrinsic,hofstra2008assess}. Broadly speaking the current understanding of the generation and maintenance of SWC is that sleep is governed by two main mechanisms involving on  regulation by 24-hour circadian system and awake-dependent homeostatic build-up of sleep pressure. Then sleep itself feeds back to regulate the circadian system and homeostatic sleep pressure build-up. Light plays a key role in synchronising and pace-making the circadian cycle or rhythm, gated by the SWC in suppressing the melatonin production in the pineal gland, regulating alertness. Apart from these, the daily social activities play also a role in driving SWC.

In general, the current research on the human SWC has focused on experiments on small groups under controlled situations \cite{orzel2010consequences,davies2014effect}, and on studies based on questionnaires \cite{roenneberg2007human,levandovski2013chronotype}. The subjectivity introduced by these approaches makes it difficult to draw general conclusions about the dynamics of the SWC especially when determining which of the possible exogenous clocks it follows. Also, it has been suggested that humans living in modern societies are subjected to new environments, like electrical lighting\cite{wright2013entrainment,de2016ancestral}, and exposure to these has disrupted and changed people's natural sleeping habits. Studies of the sleeping patterns of people living in pre-industrial societies \cite{yetish2015natural} have shown that their sleeping times are similar to those of modern societies, and that temperature could play an important role in the dynamics of sleep. In the recent past the presence of new communication technologies as well as the accessibility to large-scale techno-social datasets (`Big data') have allowed the study of human behaviour from diverse perspectives applying  reality mining techniques. In particular, mobile phone call detail records (CDRs) have been analysed to study intrinsic mental health \cite{torous2015realizing}, social networks \cite{kovanen2013temporal,eagle2009inferring,jiang2013calling}, sociobiology \cite{aledavood2015daily,david2015communication,Bhattacharya160097,aledavood2016channel}, as well as behaviour of cities \cite{louail2014mobile,sun2013understanding}. We use the CDRs to study the dynamics of daily periods of low calling activity characterised by the calls being significantly reduced or almost absent.

Users of the mobile phone network have specific time intervals when their calling activity ceases and it is expected that these periods are characterised by users taking rest. Each day shows two periods of low calling activity, the first after lunch time around 4:00pm, and the second coinciding with the night period, centered around 4:00am. In turn, these two periods delimit two regions of high activity, one peaking around noon and the second one around 8:00pm (see Fig.~\ref{Fig1}B). The daily calling activity described by these four periods follows a complex dynamics across the year and along different geographical zones, allowing it to be used to provide insight into the dynamics of human resting or sleeping pattern.

\section*{Results}
We analyse a dataset containing anonymised CDRs corresponding to a $12$ month period (in $2007$) from a mobile phone service provider having subscribers in a number of European cities. The dataset contains more than $3$ billion calls between $50$ million unique identifiers, from which $10$ million were associated with individuals having a contract with the company in question. The remaining identifiers belong to the subscribers of other companies or land-lines. Each call in the dataset involves at least one subscriber. For the majority of the subscribers, the age, gender, postal code, and location of the most accessed cell tower (MACT) are available, and we include into the analysis only those subscribers (termed as `users' from here on) whose demographic information is complete. A user is considered to ``live in a city'' if the following three geographical locations -- the associated city centre, the location of the MACT and the centre of the postal code are sufficiently close (details in SI). We choose cities having more than a hundred thousand inhabitants in the year 2007, such that our final analysis takes into account a set of $36$ cities with around 1 million users in total. 

The calling activity of an entire city depends on a number of factors, but it can be described in terms of two variables: the time of the day, and the date of the year. From the dataset,  we calculate the probability distribution $P_{all}(t,d)$ of finding an outgoing call at time $t$ of a day $d = (1,...,365)$ by a caller living in a particular city. 
 We define a `day' starting from 4:00am of a calendar day and running to 3:59am of the next calendar day. 

\begin{figure}
\centering
\includegraphics[width=.5\linewidth]{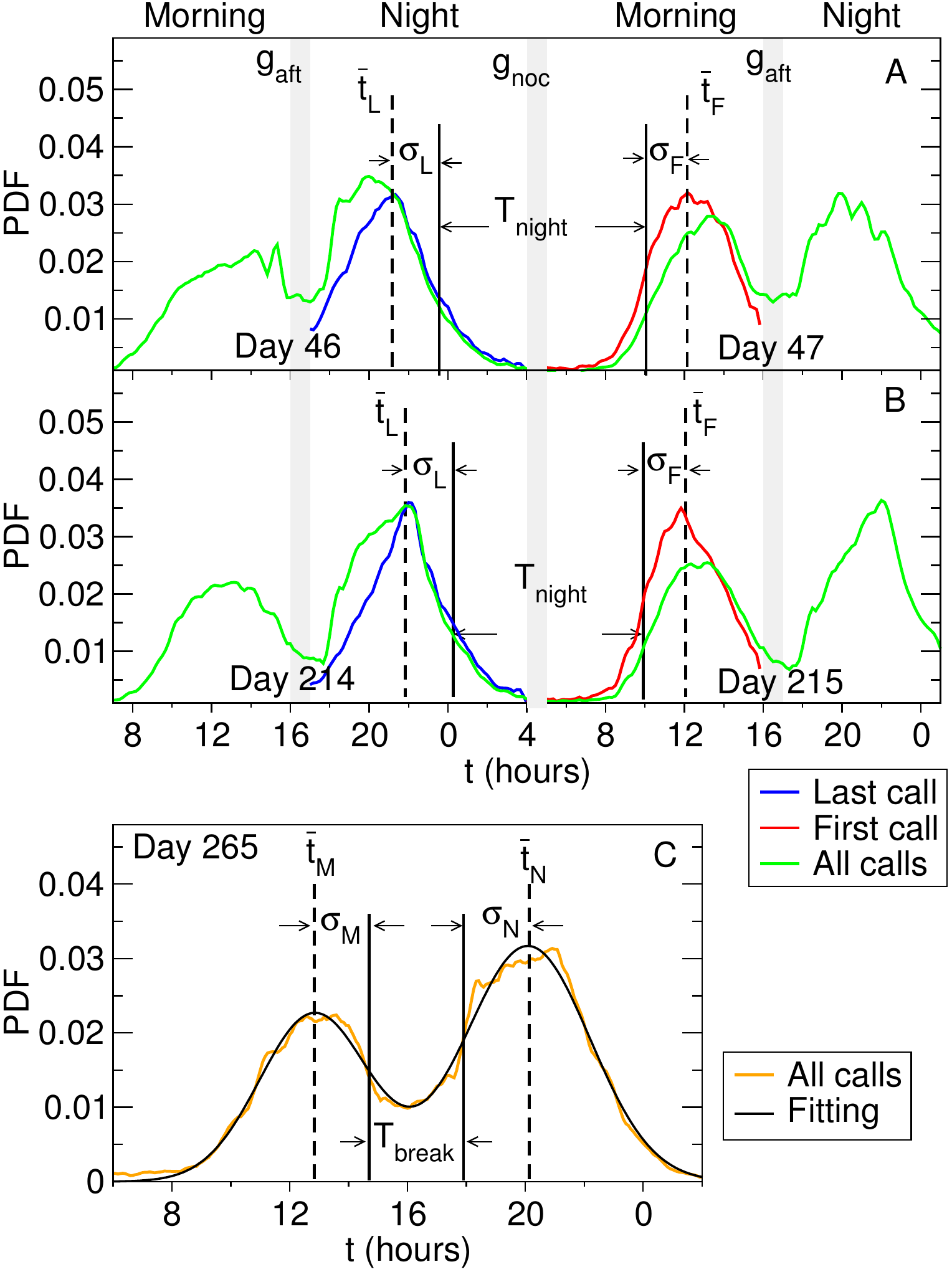}
\caption{(Top panel) 
Probability distribution functions (PDF) of outgoing calls at time $t$ of the day in a city for two sets of consecutive days of 2007: $P_{all}$ when all the outgoing calls are included (Green), $P_L$ when only the last outgoing calls in the {\it night} are included (Blue), and  $P_F$ when only the first outgoing calls in the {\it morning} are included (Red). From $P_L$ and $P_F$, their means, $\overline{t}_L$ and $\overline{t}_F$, and their standard deviations $\sigma_L$ and $\sigma_F$, respectively, are calculated and used to define the period of low calling activity (PLCA) as the region bounded by $\overline{t}_L$ and $\overline{t}_F$ and determine its width $T_{night}$ as the time interval between $\overline{t}_L+\sigma_L$ and $\overline{t}_F-\sigma_F$. For the day 46 (middle of February, set A),  $T_{night}\approx 10.5$ hours, whilst for the day 214 (early August, set B), $T_{night}\approx 9.5$ hours. (Note that $P_{all}$ is delimited by the nocturnal calling gap $g_{noc}$ (between 4:00am and 5:00am), $P_{L}$ lies between the diurnal calling  gap $g_{aft}$ (from 4:00pm to 5:00pm) and $g_{noc}$, whilst $P_{F}$ is bounded by $g_{noc}$ and $g_{aft}$). (Bottom panel) Probability distribution of outgoing calls during the day 265 (Sunday, orange line) in a city (of label 2), fitted by a superposition of two normal (Gaussian) distributions (black line), the first centered round noon and the second one in the evening (8 pm). The fitting is done with the mean values $t_M$ and $t_N$, and the standard deviations $\sigma_M$ and $\sigma_N$, corresponding to the noon and evening-centered activity modes, respectively (see details in the text). The afternoon break period $T_{break}$ is defined as $(t_N-\sigma_N)-(t_M+\sigma_M)$.
}
\label{Fig1}
\end{figure}

In Fig.~\ref{Fig1}A we show  $P_{all}(t,d)$ (green curve) during the two different pairs of consecutive days, $d$$=$$46-47$ (marking mid-February) and $d$$=$$214-215$ (the beginning of August) of 2007, for a city with around six hundred thousand inhabitants. The distribution $P_{all}(t,d)$ turns out to be  bimodal with the first mode corresponding to the calls made during the morning, peaking around noon, and the second mode being related to calls during the evening, reaching its maximum around 8:00 pm. Such bimodal patterns are found for all the days around the year and for all the cities included in this study. These two peaks are naturally delimited by two regions when the activity falls to a minimum,  one between 4:00 pm and 5:00 pm, associated mainly with the time after lunch, and a second at the end/beginning of the day,  between 4:00 am and 5:00 am, which lies inside the normal sleeping period. We will refer to the former as the afternoon calling activity minimum $g_{aft}$ and to the latter as the nocturnal calling activity minimum $g_{noc}$.

In order to study the periods of low activity, we split the day into two non-overlapping periods -- `morning' and `night', each 11 hours long, delimited by $g_{noc}$ and $g_{aft}$. Here we  define the `morning' as the time period between 5:00 am and 3:59 pm, and `night' the time period between 5:00 pm and 3:59 am on the following calendar day. Each `morning', a user can make a number of calls but to focus on the time when the calling activity starts we consider only the first call made during the `morning' for each user and construct the associated probability distribution of the time of the first call $P_F(t,d)$. Similarly, we find the last call made by every user during the `night' and construct the corresponding probability distribution for the time of the last call $P_L(t,d)$. It should be emphasized that in this study we include only the calls made by the users (outgoing calls), excluding all the incoming calls, because these may not depend on the activity pattern of the users. In Fig.~\ref{Fig1}A we compare the probability distribution for all the calls $P_{all}(t,d)$  with the corresponding distributions $P_L(t,d)$ and $P_F(t,d)$ for the times of the last call (blue) and first call (red), respectively, for two different pairs of consecutive days (during winter and summer) for the particular city with a population over six hundred thousand. The shapes of the distributions $P_L(t,d)$ and $P_F(t,d)$ depicted in Fig.~\ref{Fig1}A appear to be preserved for all the days and cities we have studied. For some special days, mainly holidays and festivities, associated the distributions where filtered out due to their atypical shape (see \ref{SI-Fig1} in SI).

\subsection{Influence of latitude in seasonal variability of low-activity period}

As regards to the calling activity presented in Fig. \ref{Fig1}A, we expect the sleeping hours to be mainly concentrated in the night period when the calling activity falls to a minimum. This nocturnal period of resting (NPR) is quantified in the following fashion. From distributions $P_L$ and $P_F$ for each city, we calculate the two means $\overline{t}_L(d)$ and $\overline{t}_F(d)$ to determine the times when the last call of the day and first call of the following morning were made, respectively. Thus the NPR is defined to be the region bounded by $\overline{t}_L(d)$ and $\overline{t}_F(d+1)$. In order to estimate the duration of the period when the calling activity has practically ceased, we determine the width of NPR, $T_{night}$, by taking into account the widths of the distributions $P_F$ and $P_L$, determined by the standard deviations $\sigma_F$ and $\sigma_L$, respectively. Hence, we can write $T_{night}(d) = 24-\left(\overline{t}_L(d) +\sigma_L\right)+\left(\overline{t}_F(d+1)-\sigma_F\right)$ (see Fig.~\ref{Fig1}A). This definition for the width of NPR does not require any introduction of arbitrary cut-off parameters. 

We have calculated $P_L$ and $P_F$ for 12 cities lying within one of three latitudinal bands, centered at  $37 ^ \circ $N, $40 ^ \circ $N, and $42.5 ^ \circ $N, in such a way that each band contains 4 cities. In Fig.~\ref{Fig4} we show the yearly variation of $T_{night}$ for these 12 cities. The plots show that the width $T_{night}$ changes across the year, being longest near the winter solstice and shortest six months later, near the summer solstice. Comparing the different plots corresponding to $T_{night}$, the following two characteristics should be noted. First, cities lying in the same latitude band show similar $T_{night}$ curves, differing only by a vertical offset. This implies that although each city has a characteristic period of low activity, its yearly variation is very similar to that in other cities at the same latitude. Secondly, the difference between the highest (near winter solstice) and lowest values (near summer solstice) of $T_{night}$ changes from latitude to latitude, being larger in southern cities and showing that there is an external factor that influences $T_{night}$ with different intensity at different latitudes. 

\begin{figure}[h]
\centering
\includegraphics[width=.75\linewidth]{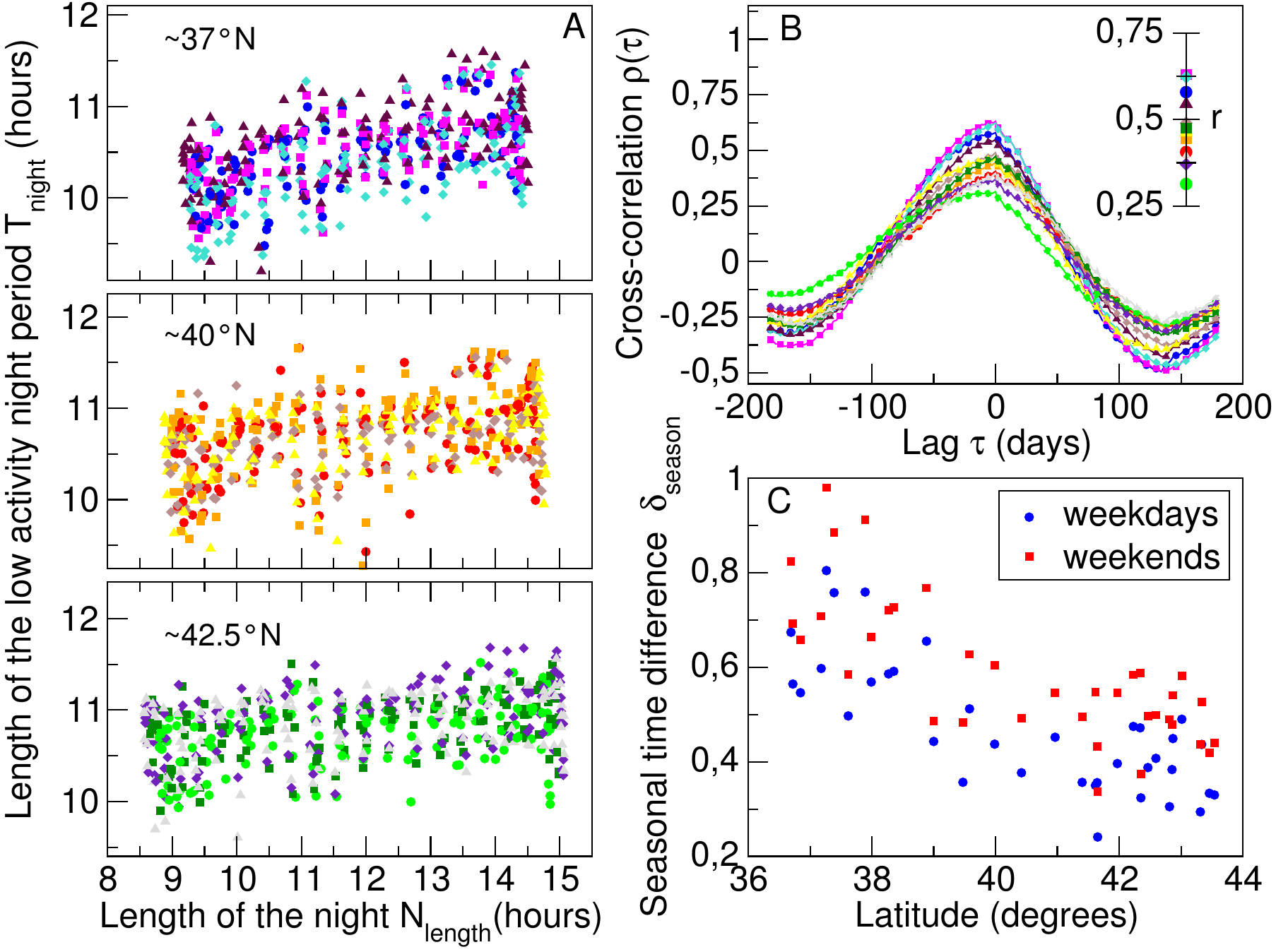}
\caption{Relation between the period of low activity $T_{night}$ and the length of the night $N_{length}$: (Panel A) Scatter plots of $T_{night}$ against $N_{length}$ for 12 cities located at one of the three different latitudinal bands $37^\circ$N (blue, magenta, turquoise and maroon), $40^\circ$N (red, orange, brown and yellow), and $42.5^\circ$N (green, dark green, indigo and gray), such that in each set the points represent the pairs ($N_{length},T_{night}$) across the 365 days of 2007 (only the Tuesdays, Saturdays and Sundays are shown). 
(Panel B) The cross-correlation $r$ between $T_{night}$ and $N_{length}$ as a function of the time lag $\tau$. Across the year $T_{night}$ and $N_{length}$ time series are synchronized, as their cross-correlation $\rho(\tau)$ reaches its maximum for the time lag $\tau=0$, thus $T_{night}$ and $N_{length}$ are in phase and as depicted in the inset, the correlation between them is for all cases very high, i.e. $r > 0.3$, $p<0.001$. (Panel C)  Latitudinal dependence of the net change $\delta_{season}$  between $\overline{T}_{night}$ at winter solstice (maximum) and at summer solstice (minimum). For cities in the south, this difference is larger than for the cities in the north. Each point corresponds to one of the 36 studied cities calculated as averages of $\overline{T}_{night}$ for weekdays from Mondays to Thursdays (blue) and for weekends from Fridays to Sundays (red). 
}
\label{Fig2}
\end{figure}

As a matter of fact, the shape of $T_{night}(d)$ throughout the year resembles the yearly behaviour of the length of the night $N_{length}(d)$, defined here as the time between the sunset and the sunrise for each city. The value of $N_{length}$ peaks around the winter solstice ($d=356$) and decreases monotonically until it reaches a minimum value around the summer solstice ($d=172$), coinciding with the behaviour of $T_{night}$. In the scatter plots of $\overline{T}_{night}$ vs. $\overline{N}_{length}$   (Fig.~\ref{Fig2}A) we can see a noticeably linear dependence between these quantities, with the southern cities ($\approx 37^\circ$N) showing the strongest effect. We use a linear regression, $\overline{T}_{night} = \beta \overline{N}_{length} + \alpha$, to quantify the yearly variation of $\overline{T}_{night}$ as a function of $\overline{N}_{length}$. Here $\overline{T}_{night}$ is defined as the average value of $T_{night}$ from Monday to Thursday each week, to characterize typical weekdays and to reduce the fluctuations of $T_{night}$. We carry out the regression for $36$ different cities (including the above-mentioned 12 cities) located between the $36 ^ \circ $N and $44 ^ \circ $N, for weekdays. Similarly, to characterize typical weekend days, we have calculated $\overline{T}_{night}$ averaged across  Fridays, Saturdays and Sundays. In order to study the effect of latitude on the seasonal variation of the NPR, we have calculated the net change $\delta_{season}$  between $\overline{T}_{night}$ for the winter solstice (maximum) and for the summer solstice (minimum), given by $\delta_{season}=\beta \left(\overline{N}_{length}(d=356)-\overline{N}_{length}(d=172)\right)$. This simple definition provides a way to calculate how the latitude affects the range where the period of low activity varies. For the 12 cities studied, we find that ${N}_{length}$ and ${T}_{night}$ are highly correlated, as can be seen in the inset in Fig.~\ref{Fig2}B (in general $r\approx 0.5$ and all $p$ values with more than $99\%$ confidence level). 
Besides this high correlation between ${N}_{length}$ and ${T}_{night}$, we explore if there is a time lag between their corresponding time series. For this, we calculate the cross-correlation $\rho$ between the time series for the same 12 cities as above. In Fig.~\ref{Fig2}B it can be seen that $\rho$ reaches its maximum when the time lag $\tau$ is zero, meaning that these time series vary in the same phase. In Fig.~\ref{Fig2}C we show $\delta_{season}$ for each of the 36 cities at their own latitudes. Remarkably, the time difference $\delta_{season}$ between winter and summer is always at least 15 minutes larger for the southern cities compared to the northern cities. For weekends we observed a difference of 30 minutes, with the ratio close to 2 (30 minutes in the northernmost cities against 1 hour in the southernmost ones).  The slopes of the linear regression obtained for each one of the 36 cities can be seen in the SI, Fig.~\ref{SI-Fig2}.
 Cities around the same latitude have similar slopes ($\beta$s), and these slopes are larger for cities at the southern latitudes. 

\subsection{Ambient temperature effects on the duration of diurnal resting period}

For all the 36 cities included in this study and for almost every day, $P_{all}$ shows apart from the nocturnal period of resting or predominantly night-time sleeping period, another low activity period centered around the diurnal gap $g_{aft}$, ranging from 4:00 pm to 5:00 pm (Fig.\ref{Fig1}C). The dynamics of this diurnal afternoon period of resting  (DPR), associated with a decrease in human activity, can also be traced from the $P_{all}$ dynamics though here the decrease is less dramatic than in the nocturnal case. This is due to the fact that $P_{all}$ can be satisfactorily described by the sum of two normal distributions, with one of the activity peaks centered around the noon, and the other in the evening around 8:00 pm as seen in \ref{Fig1}C. Unlike in the NPR case where almost all the activity of people ceases due to them sleeping, the DPR shows a more moderate decrease in activity between these two peaks. This is associated with only a fraction of people having naps or taking rest. The length of this low activity period DPR can be measured with the fitting function given by $F(x)=\frac{a_0}{\sigma_M\sqrt{2\pi}} e^{-0.5\left((t-\overline{t}_M)/\sigma_M\right)^2}+\frac{a_N}{\sigma_N\sqrt{2\pi}} e^{-0.5\left((t-\overline{t}_N)/\sigma_N\right)^2}$. Here $\overline{t}_M$ and $\sigma_M$ are the mean and the standard deviation of the noon-centered distribution and $\overline{t}_M$ and $\sigma_M$ the corresponding values for the evening-centered one. Hence we define the duration of the DPR or the afternoon break period to be  
$T_{break}$ as $(\overline{t}_N-\sigma_N)-(\overline{t}_M+\sigma_M)$ (Fig.\ref{Fig1}C). The yearly change in $T_{break}$ for 12 cities located at one of three different latitudinal bands $37^\circ$N, $40^\circ$N, or $42.5^\circ$N is depicted in Fig.~\ref{Fig4}. 
\begin{figure}[h]
\centering 
\includegraphics[width=0.5\linewidth]{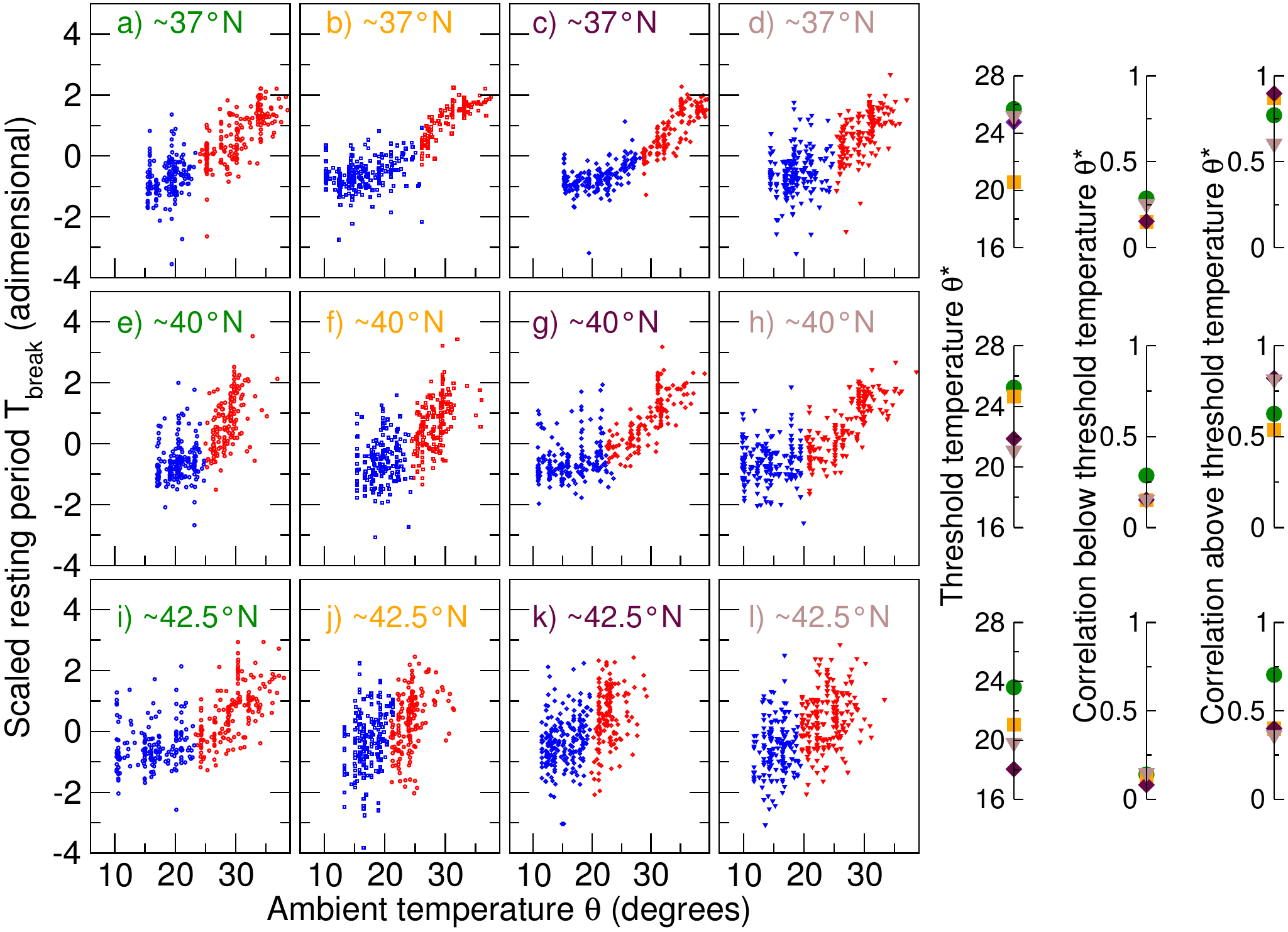}
\caption{Relation between the afternoon break period $T_{break}$ and the maximum daily temperature $\theta$, for 12 different cities located at one of the three different latitudinal bands: (top) $37^\circ$N, (middle) $40^\circ$N, and (bottom) $42.5^\circ$N. (Left panels) For each city (\textsl{a} to \textsl{l}), a point represents a pair ($T_{break},\theta$) for one of the 365 days of 2007. The scatter plots show that $T_{break}$ behaves differently for different temperature ranges (blue points correspond to ``cold'' or ``cool'' days and red points to ``warm'' or``hot'' days), separated by a threshold temperature $\theta^*$  below which $T_{break}$ seems to stay more or less constant and above it increasing with temperature. (Right panel) For each of the three latitudinal bands variabilities of: the threshold temperature $\theta^*$ (first column) and the correlation $r$ between $T_{break}$ and $\theta$ when the temperature is below (second column)  and above (third  column) the threshold temperature $\theta^*$. (The color labeling (green, orange, indigo or brown) of cities in each latitudinal band is the same in both panels).  
The correlation are calculated using points above the temperature thresholds, turning out to be strong in all cases with values $>99\%$. For temperatures below the threshold, the correlation is weak with $p$-values in general greater than $0.05$. The points were classified as below (blue) and above (red) the threshold by using spectral clustering algorithm.
}
\label{Fig3}
\end{figure}

In order to study the influence of the daily temperature on the dynamics of $T_{break}$, we present it as a scatter plot with the maximum daily temperature $\theta$ for the 12 above mentioned cities, as depicted in Fig.~\ref{Fig3}. It can be seen that $T_{break}$ follows two different dynamics, distinguishable by a threshold temperature $\theta^*$, below which $T_{break}$ seems not to be strongly influenced by the temperature and above which there is a clear linear trend (see Fig.~\ref{SI-Fig3} in the SI for a comparison between the time series).  To make this distinction more visible and to estimate the threshold, we apply a classification algorithm\cite{shi2000normalized} to cluster the points into two sets separated by the threshold temperature $\theta^*$ that maximizes the difference between sets and minimizes the difference inside each set. The results for each city can be seen in Fig. \ref{Fig3}, in the form of two sets (red and blue points), with a threshold temperature ranging from $18^\circ$ to $25^\circ$ (Fig.~\ref{Fig3}). Once each set is split into two subsets, we calculate the correlation between $T_{break}$ and $\theta$ for the set of cities (Fig.~\ref{Fig3}, right panels). We observe a weak correlation when the temperature is below the threshold, i.e. for almost all of the 12 cities $r\leq 0.25$ and $p\approx 0.05$. On the other hand, when the temperature is above the threshold, we observe the correlations to be strong, i.e. $r\geq 0.5$ and $p\approx 0.0001$, which indicates a strong influence of ambient temperature on the afternoon break period $T_{break}$.

Previously we have shown that both, the daily NPR and DPR (i.e. the nocturnal and diurnal resting periods), as well as the characteristic times $T_{night}$ and $T_{break}$ follow specific dynamics showing seasonal and geographical variations. As depicted in Fig. \ref{Fig4} these resting periods seem to follow opposite dynamics across the year such that when added up, the resulting total daily resting period shows only a small variation over the year (see Fig. \ref{SI-Fig4} in the SI). This lends support to the notion that the total daily resting time of people is more or less conserved.

\begin{figure}[t]
\centering
\includegraphics[width=0.5\linewidth]{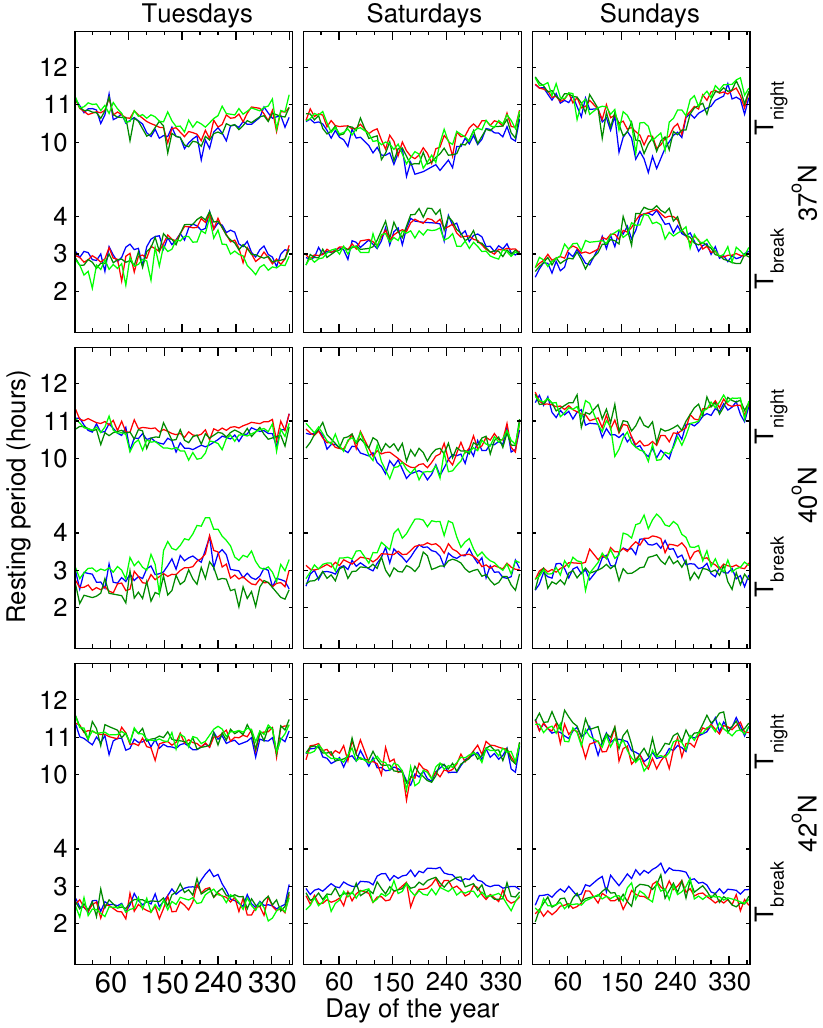}
\caption{Periods of low calling activity or of resting: $T_{break}$ and $T_{night}$ for 12 different cities across 2007 for 3 different days of the week. The cities are located in one of the three different latitudinal bands (top) $37^\circ$N,(middle) $40^\circ$N, and (bottom) $42.5^\circ$N. 
Here $T_{break}$ and $T_{night}$ characterize the diurnal and nocturnal resting times, which show seasonal variations that appear to counterbalance each other.
}
\label{Fig4}
\end{figure}



\section*{Discussion}

At the individual level the SWC is known to depend on various physiological and social factors that make it difficult to identify the precise causes influencing its dynamics. However, studying the yearly variation of the NPR averaged over a big population gives us quite unique insight into its complex behaviour and helps us to identify some of the key influencing factors. The SWC is bounded inside the NPR and we expect that both entities follow similar dynamics but with somewhat different onset and termination times. We have shown that the duration of the NPR, characterized by the width $T_{night}$, closely follows the seasonal variation in the duration of the night. Moreover, we have observed that the duration of NPR is influenced by the latitude of the city in question. The difference between $T_{night}$ in winter and in summer for the people in cities located at $36^\circ$N was found to be almost twice as much as that for the people in cities located $8^\circ$ further north. It seems that the activity of people in southern cities are most affected by the length of the night (or conversely length of the daylight), despite the fact that in northern locations the length of the night changes faster and spans a bigger time interval than more southern latitudes.

Human physiology and its hormonal regulation is expected to govern the dynamics of the SWC at the individual level. In particular, the melatonin hormone is known to follow a circadian rhythm \cite{arendt2000melatonin} and has been linked to the SWC. For individuals, melatonin secretion commences during the evening between 8 pm and 10 pm, peaking between 2:00 am and 4:00 am. However, the onset and the duration of melatonin secretion during the circadian cycle is in turn related with the exposure to light and darkness, and considerable research has been done \cite{brown1994light,dawson1993melatonin,broadway1987bright,vondravsova1997exposure} showing how changes in the onset and length of the light exposure disrupts the secretion of melatonin, as well as the SWC \cite{wehr1998effect}. For people living at latitudes far away from the equator, where the seasonal changes in the onset and length of daylight are large, it has been shown that their melatonin cycle is seasonally altered and this perturbation gives rise to disruptions in their SWC \cite{wehr1991durations}. Thus it could be possible that the melatonin cycle is the biological intermediary entraining the NPR (and the SWC) with the daylight duration along the year. 

On the other hand, the diurnal period of resting DPR is found to depend rather strongly on the temperature, making its duration during hot days longer, but showing no noticeable variation during cold or cool days.  Surprisingly, however, this increase in the daily diurnal resting time is found to be counterbalanced by the decrease of the daily nocturnal resting time. Hence it seems that on warm or hot days a bigger fraction of the total resting period is taken by the afternoon break, thereby reducing the homeostatic pressure for sleeping and consequently the need for nocturnal resting period. 

The evidence of the trade-off between the afternoon and nighttime rest or sleep at latitudes closer to the tropics has implications for our understanding of the time budgets of tropical hunter-gatherers who have also been reported to take afternoon `naps' for about a quarter of a day  
\cite{yetish2015natural}. Our results suggest that, in high temperatures tropical habitats, humans find it difficult to remain active during the afternoon when ambient temperatures are at their highest, much as in the case of for most monkeys and apes\cite{korstjens2010resting,david2016bipedality}. Fig.~\ref{Fig3} suggests that there is a critical threshold at which rest becomes necessary and that this occurs at ambient temperatures somewhere between $20-25^\circ$C (though `naps' of significant length are probably not common until ambient temperatures exceed $\approx 30^\circ$). This has important implications in that it removes a significant amount of time from the active working day, thereby reducing the time that can be devoted to ecologically or economically more important activities. Such constrains have imposed significant limitations on our species' ability to evolve complex societies \cite{dunbar2014human}.

\section*{Methods}
\subsection*{Time binning and smoothing of distributions}
To generate each probability distributions, we divide the temporal axis in 5 minutes bins, in such a way that each $P_{all}$ contains 268 points, and $P_L$ and $P_F$ 132 points each one. In some cases, mainly for cities with small population, their associated probability distributions show fluctuations due to small sample size, then we apply a smoothing process to each distribution in order to reduce the noise, using the Savitsky-Golay \cite{savitzky1964smoothing} algorithm (7 points and degree 4).
\subsection*{Filtering}

\subsubsection*{Geographical location determination}
From our dataset, two locations could be associated to each user. First one is the location of the most accessed cell tower (MACT) by the mobile device. From the zip code of the domicile of the user we get the second one, namely the location of the centre of the postal zone (postal location). These two locations were used when determining which users  should be included in the analysis. We discarded those users with at least one of these two locations missing.

To reduce the noise due to poor sampling, we chose from the whole set of towns included in the mobile phone network only those cities with more than hundred thousand inhabitants in the studied year, and lying between latitudes $36^\circ$N and $44^\circ$N,  which accounts for 36 cities. For each selected city, we encircled it with a 30 km diameter circle, and use the center of the circle as the associated geographical location of the city. All the studied cities fit inside a $30$ km diameter circle.

Finally, we choose only those users who live in one of these cities. To ensure this into some extent, we impose the next two restrictions:
\begin{itemize}
\item the distance between the city location and at least one of the associated user locations (MACT and postal) should be less that $15$ km, and
\item the distance between the MACT and postal locations of the user must be less than $30$km.
\end{itemize}

There is no way to verify that the user actually lives in the associated city, but one could expect that almost every user lives in the same city as the one specified by the domicile, and that the MACT is indicative of their usual location. Thus the imposed restrictions seem to be a good way to ensure that a big fraction of the included users are correctly assigned to their actual city.

From the demographic information available, one can find the age of the  users ranges from eighteen to more than hundred years. In this work we choose only users with age between 30 and 75 years old, because the calling pattern in younger people (mainly in the 18 to 25 years range) is more erratic than for older people. Older people (75+) were excluded the analysis due to the sparse calling activity that many of them had in the CDRs. After the previous filtering, the included users into the analysis were $925,135$ individuals.
\subsection*{Data availability}
The datasets analysed during the current study are not publicly available due to a signed non-disclosure agreement. The dataset contains sensitive information of the subscribers, particularly age, gender, postal address and the location of the most accessed cell tower. Calling time distributions from which all the results shown in this work were generated are available from the last author on reasonable request.

\bibliography{sample}

\section*{Acknowledgements}
AG and KK acknowledge project COSDYN, Academy of Finland (Project No. 276439) for financial support. DM, KB, and KK acknowledge EU HORIZON 2020 FET Open RIA project (IBSEN) No. 662725. DM also acknowledges CONACYT (Mexico), grant No. 383907. RD acknowledges European Research Council for the Advanced Investigator Grant No. 295663. Authors thank AL Barab\'asi for the dataset used in this research.


\section*{Additional information}

\subsection*{Competing financial interests} 
The authors declare no competing financial interests.
\section*{Author contributions statement}
K.B., A.G. and D.M. carried out the analysis of the data. All the authors were involved in designing the project and the preparation of the manuscript.
\clearpage
\appendix

\section*{Supplementary Information}
\subsection*{Time series filtering due to holidays and atypical days filtering, and mean times for the first and last call
}

When calculating the probability distributions from the CDRs, we found that for some particular days and cities, the behaviour of these distributions were atypical. The corresponding days when these anomalies appeared were in general special days, when some festivity or holiday was held, mainly Easter holidays (days 90 to 98) and the days around Christmas and New Year. In order to avoid possible fluctuations introduced by these atypical days, we  excluded these days from the analysis. When there was a national holiday, the point corresponding to that day was removed from the time series of all cities, and if the atypical day was a local festival or local event,  that day was removed only for that city. Examples of this filtering can be seen in  Fig.~\ref{SI-Fig1}.

\renewcommand{\thefigure}{A\arabic{figure}}

\setcounter{figure}{0}

\begin{figure}[h]
\centering
\includegraphics[width=0.5\linewidth]{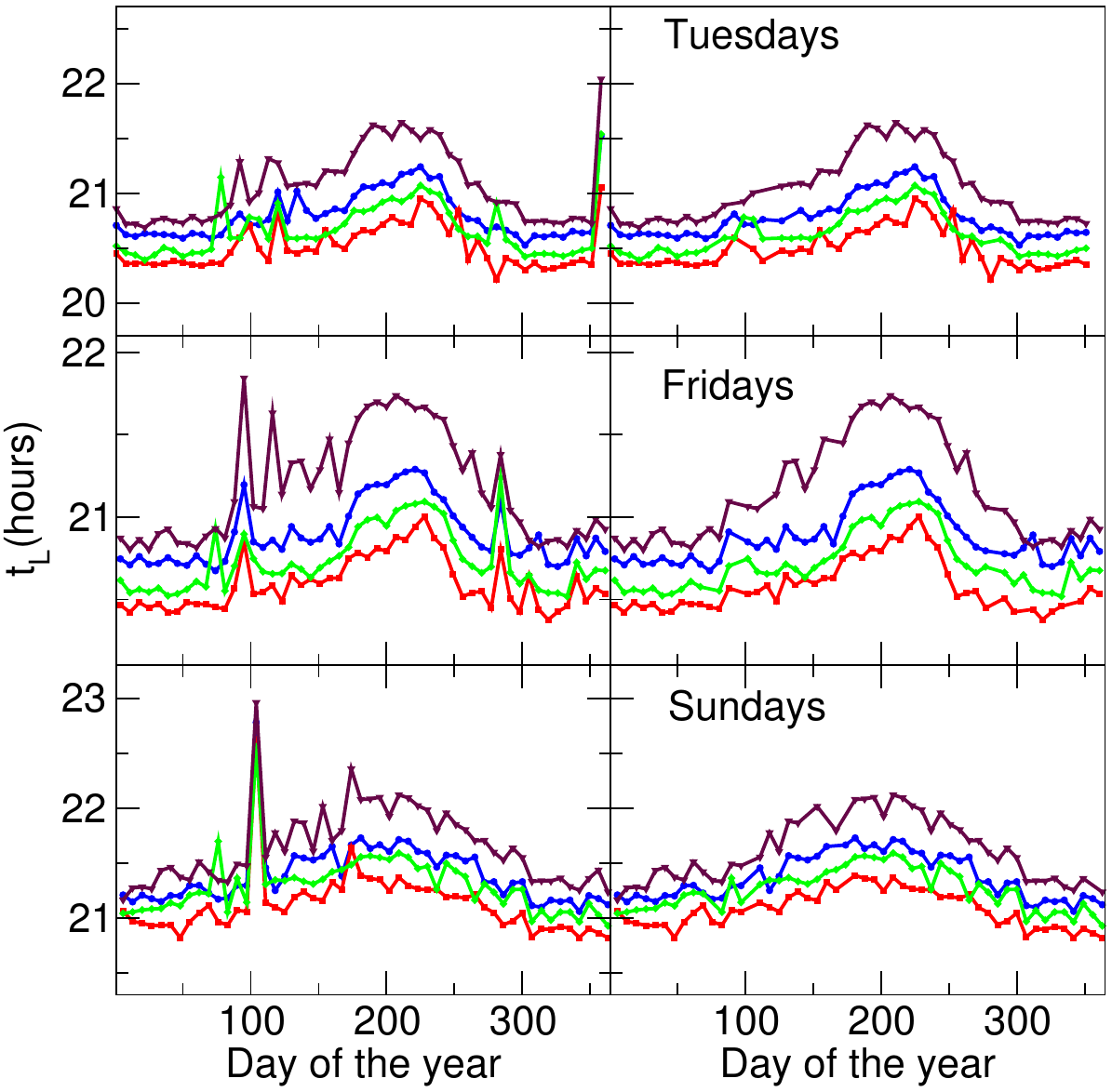}
\caption{Removal of days when a national holiday, local festival or special event is present for the most 4 populated cities. (left column) Original $t_L$ for each city. (right column) Filtered $t_L$.}
\label{SI-Fig1}
\end{figure} 

\begin{figure}[t]
\centering
\includegraphics[width=0.5\linewidth]{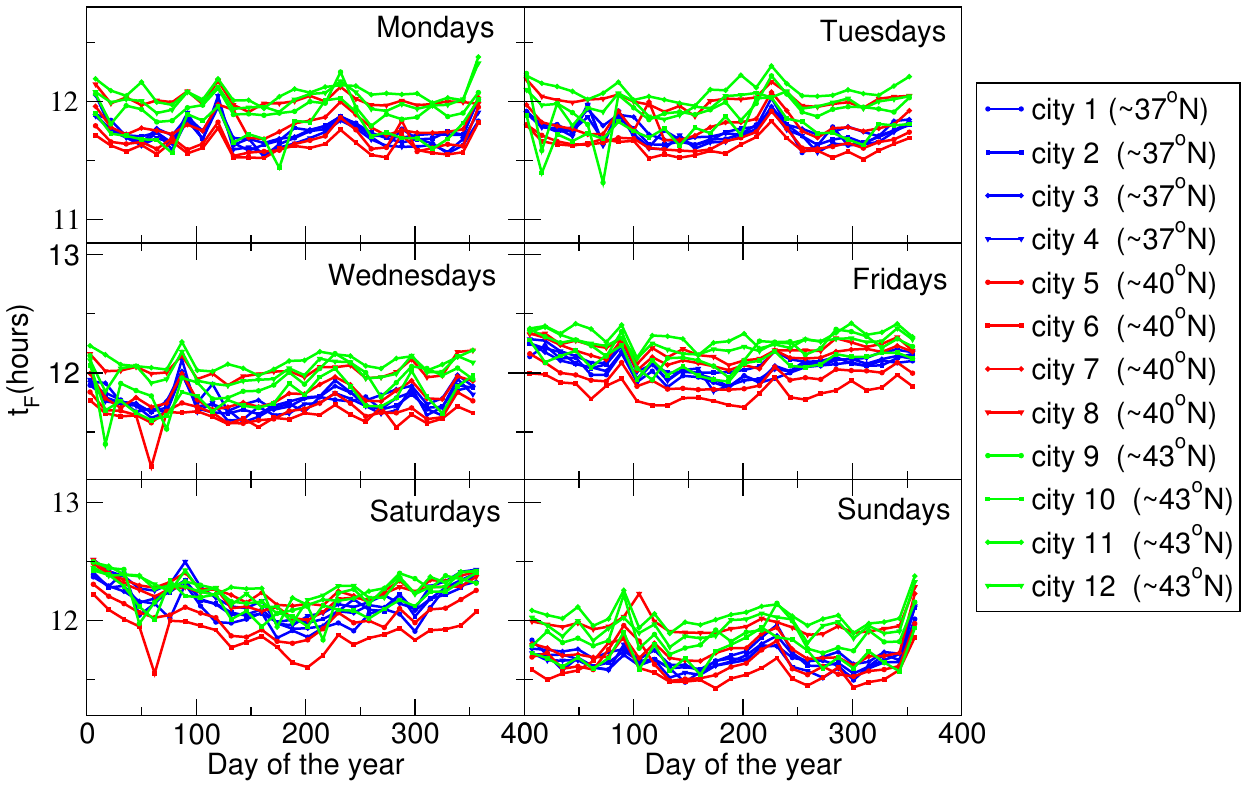}
\caption{Mean time between of the first call $\overline{t}_F$ for 12 different cities during one year. The cities are  grouped in 3 sets with 4 cities per set. Each set is located along one of these three latitudes $37^\circ N$, $40^\circ N$, or $43^\circ N$.  The time when the first call is made slightly changes along the year, specially during the weekends. Also southern cities start the calling activity slightly earlier.  The peaks shown in the graphs correspond to certain special days, like holidays and festivals. Each time series is vertically shifted by an amount proportional to the time difference between the local sun transit time of each city and the corresponding time at some reference point in the middle of the band, to enhanced the visualization of the time series in each band, which show similar shapes.}
\label{SI.Fig3}
\end{figure}

\begin{figure}[t]
\centering
\includegraphics[width=0.5\linewidth]{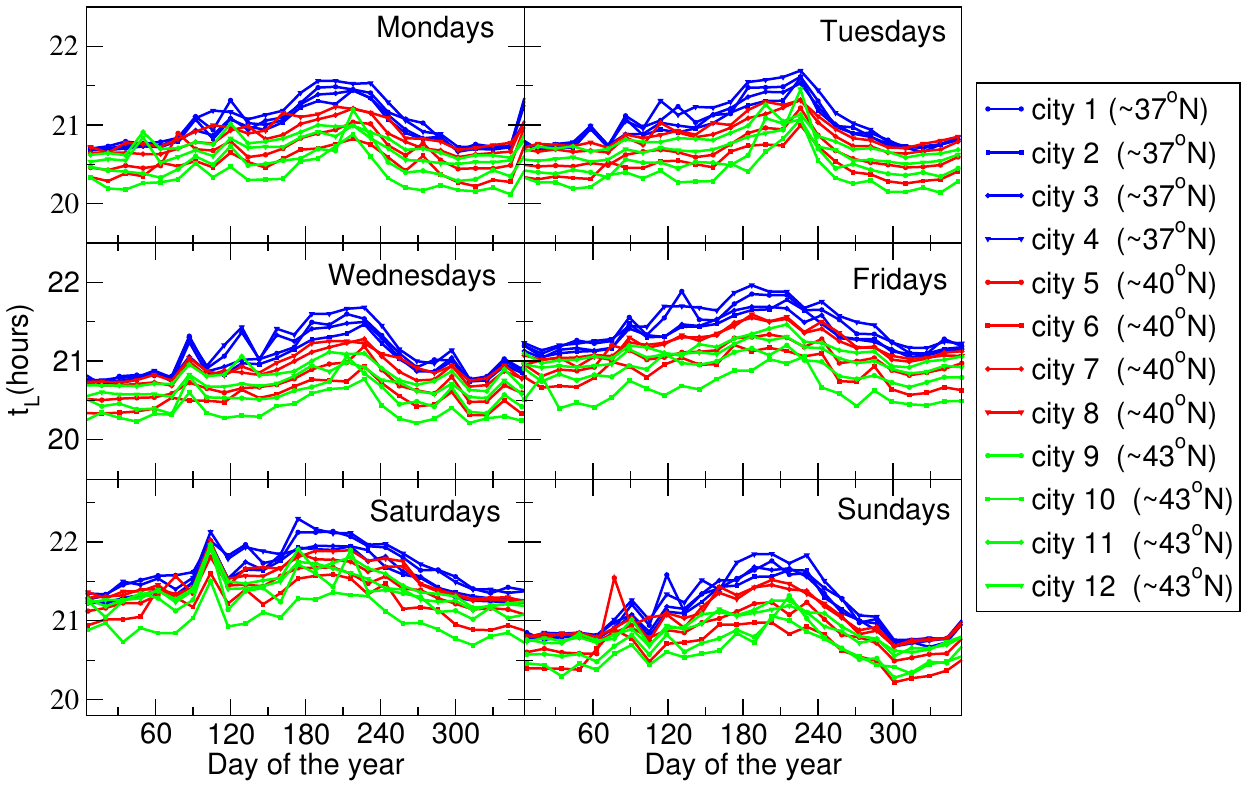}
\caption{Mean time of the last call $\overline{t}_L$ for 12 different cities during one year. The cities are  grouped in 3 sets with 4 cities per set. Each set is located along one of these three latitudes $37^\circ N$, $40^\circ N$, or $43^\circ N$.  The time when the last call changes strongly   along the year, specially during the weekends. Southern cities cease the calling activity later, and the size  of the seasonal change is stronger than in the case of northern cities (the time difference between summer and winter days is bigger for southern cities). The peaks shown in the graphs correspond to certain special days, like holidays and festivals. Each time series is vertically shifted by an amount proportional to the time difference between the local sun transit time of each city and the corresponding time at some reference point in the middle of the band, to enhanced the visualization of the time series in each band, which show similar shapes.}
\label{SI.Fig4}
\end{figure}

In Fig. \ref{SI.Fig3} and  Fig. \ref{SI.Fig4} the mean times for the first call $\overline{t}_L$ and for the last call $\overline{t}_F$ are shown, respectively, for 12 cities located along three different latitudes. To enhance the visualization of the time series in each band so as to emphasise the similarity in shape, each time series is vertically shifted by an amount proportional to the time difference between the local sun transit time of each city and the corresponding time at some reference point in the middle of the band.  

\subsection*{Linear regression of $T_{night}$ on $N_{length}$}
From the linear regression, $\overline{T}_{night} = \beta \overline{N}_{length} + \alpha$, used quantify the yearly variation of $\overline{T}_{night}$ as a function of $\overline{N}_{length}$, in subsection ``Influence of latitude in seasonal variability of low-activity period'' in the main text, only $\delta_{season}$ was shown. Here we present the corresponding plot of the slopes $\beta$ of the linear regression. Again, $\overline{T}_{night}$ is defined as the average value of $T_{night}$ from Monday to Thursday each week, to characterize typical weekdays and to reduce the fluctuations of $T_{night}$.

\begin{figure}[h]
\centering
\includegraphics[width=0.5\linewidth]{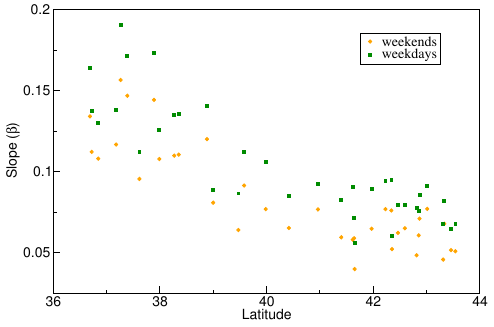}
\caption{Latitudinal dependence of the net change $\delta_{season}$  between $\overline{T}_{night}$ at winter solstice (maximum) and at summer solstices. Coefficient of linear regression, $\beta$, obtained for each of the 36 studied cities as a function of its latitude. In southern cities, the width of the NPR changes faster (larger $\beta$s) and the difference between their minimum (summer solstice) and maximum (winter solstice) values is larger than in case of the cities in the north. Weekdays points (blue) were calculated from the average $\overline{T}_{night}$ from Mondays to Thursdays. For weekends points (red), Friday to Sunday were used to calculate the average $\overline{T}_{night}$.}
\label{SI-Fig2}
\end{figure}

\subsection*{Ambient temperature and afternoon break period time series}

\begin{figure}[t]
\centering
\includegraphics[width=0.5\textwidth]{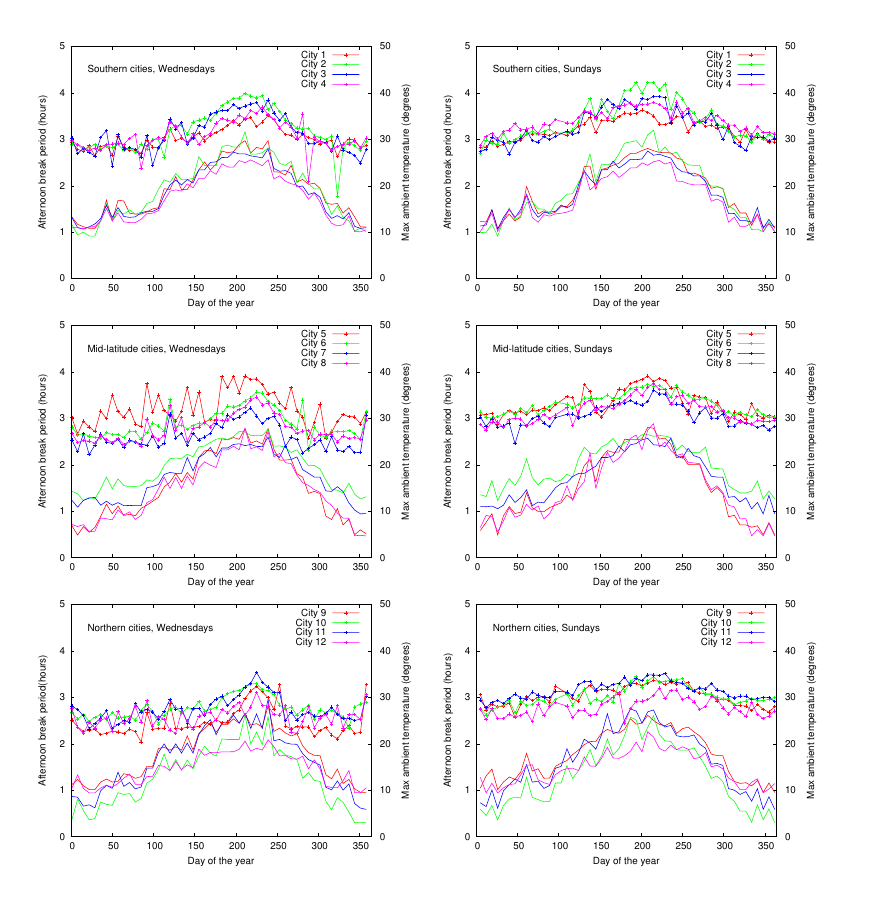}
\caption{Afternoon break period yearly dynamics compared with the corresponding temperature time series, for 4 different cities lying at different $37^\circ$N,  $40^\circ$N and  $42.5^\circ$N latitude. The lines with markers represent the length of the afternoon break resting along the year, whilst those without markers represent the max temperature  in those cities. Each city is represented by the same color in both time series. Wednesdays and Sundays are shown }
\label{SI-Fig3}
\end{figure}

Ambient temperature $\theta$ series during 2007 were obtained from the databases available from the national meteorological institutes of the countries where the cities are located. The available information contains the daily minimum, mean and maximum ambient temperatures on one of the monitoring station located in each city. For some cities, the information for some day was not present in the dataset, and the missing data was interpolated after applying a smoothing process (Savitzky-Golay) to the time series. We compare $\theta$ and afternoon break period $T_{break}$ for 12 cities, located along one of these three latitudes $37^\circ N$, $40^\circ N$, or $43^\circ N$, and a there is a strong similarity between their behaviour  during warmer months (April - October generally). From the beginning of that period, $\theta$ and $T_{break}$ increase monotonically, reaching their maximum value around August and then monotonically decrease until beginning of Autumn. Outside that period (colder months), $\theta$ follows its seasonal variation, but $T_{break}$ stagnates, and shows no relation with the ambient temperature, as can be seen in \ref{SI-Fig3}.

\subsection*{Total daily resting period}
The total period of low activity or of resting defined as $T_{rest}=T_{break}+T_{night}$ is calculated. $T_{rest}$ is the consequence of two competing processes, the afternoon or diurnal resting period $T_{break}$ and the night or nocturnal resting period $T_{night}$. Despite the seasonal variation of these two processes, the yearly changes $T_{rest}$ are smaller and, particularly during the weekends appears to be uniform. 
\begin{figure}[t]
\centering
\includegraphics[width=0.5\linewidth]{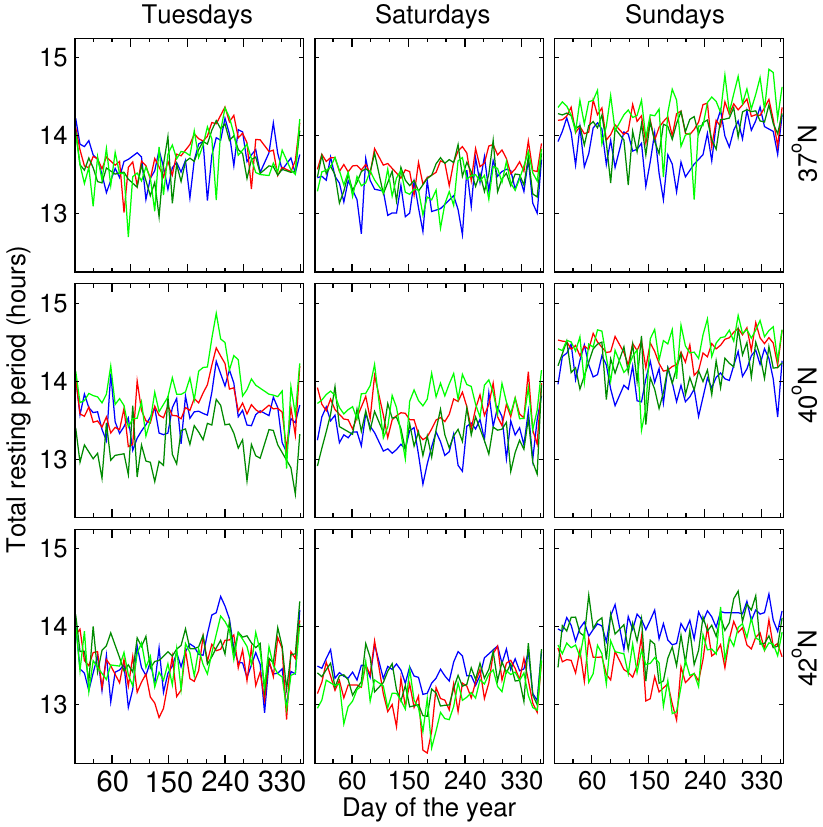}
\caption{Total period of low calling activity. The total period of low activity or of resting defined as the sum of $T_{break}$ and $T_{night}$ for 12 different cities across 2007 for 3 different days of the week. The cities are located in one of the three different latitudinal bands (top) $37^\circ$N,(middle) $40^\circ$N, and (bottom) $42.5^\circ$N. $T_{rest}$ is the consequence of two competing processes, the afternoon or diurnal resting period $T_{break}$ and the night or nocturnal resting period $T_{night}$. Despite the seasonal variation of these two processes, $T_{rest}$ is almost uniform across the year, particularly during the weekends.
}
\label{SI-Fig4}
\end{figure}
 
\subsection*{Estimation of the fraction of users active during afternoon break}

Despite the noticeable reduction of the calling activity during the afternoon period on any given day, there is a set of users calling during that time. This prevents the calling activity from falling to zero unlike in the case of the night resting period. There are multiple reasons for this background activity.  For example, some users take an afternoon break in their routine activities before or after the characteristic time of the population as a whole, while some individuals simply do not take this break. 
We make a gross estimation of the fraction of inactive users at that time in the following fashion. We focus on subset of `active' users, including in it only those who have at least a certain number of calls during the day and making their calls in both halves of the day, namely, the morning and the evening periods. With these criteria, we can expect that the included users have enough calls that are spread over the entire span of the day. Once we have determined this set of users, we divided the day into three different time periods between 4:00 am to 2:00 pm, 2:00 pm to 5:00 pm (approximately coinciding with the $T_{break}$ period), and 5:00 pm to 4:00 am (of the next calendar day), respectively.

\begin{figure}[t]
\centering
\includegraphics[width=0.5\linewidth]{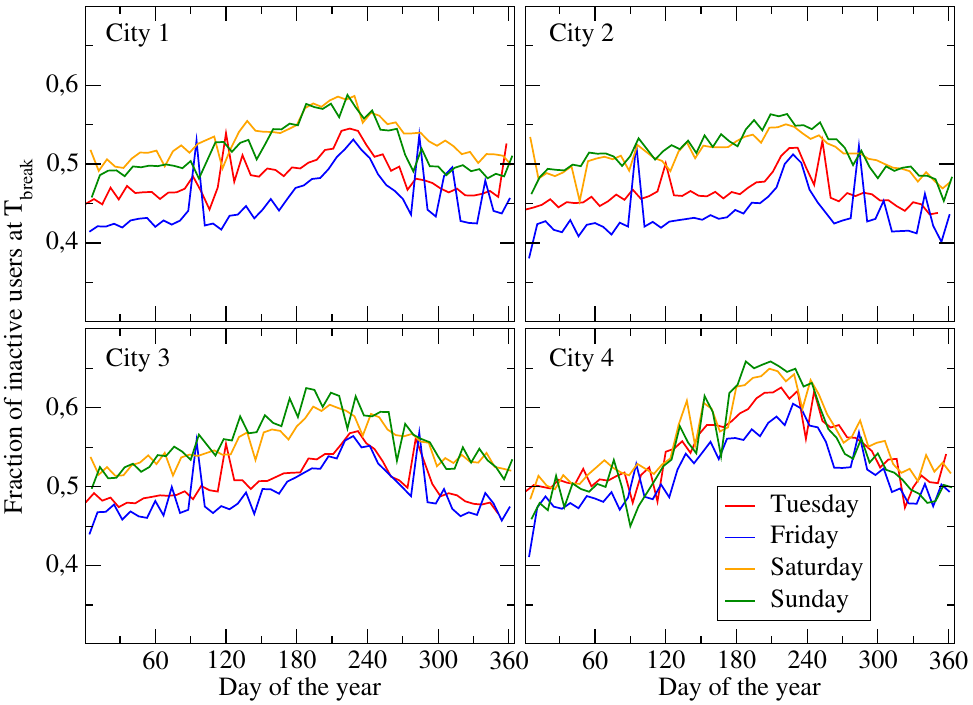}
\caption{Fraction of inactive users during afternoon resting period. The set of users used in this analysis made more than three calls that day, at least one before and one after 3:30 pm. From these users, we calculate the fraction of them that were inactive during the period 2:00pm - 5:00 pm. The fraction was calculated for the four most populated cities across 2007 for 4 different days of the week. Cities 1, 2, 3 and 4 are located around latitudes $40^\circ$N, $42^\circ$N, , $39^\circ$N, and $37^\circ$N, respectively.
}
\label{SI-Fig5}
\end{figure}

To estimate the fraction of users resting during $T_{break}$ period, we include those users with $3$ or more calls over the day (with other restrictions  as stated above), and calculate the fraction of users that don't make calls during the $T_{break}$ period. This coarse estimation gives us insight of how active users behave along the day, and allows us to determine how many of them reduce their activity during the afternoon break. The subset of users are at best a sample of the population, but at least represent the overall behaviour. The results for the four most populated cities are shown in Fig. \ref{SI-Fig5} and it can be seen that the fraction of inactive users during the afternoon break is around $0.5$ across the year.

\end{document}